\documentclass[aps,prd,showpacs,preprintnumbers,amsmath,amssymb]{revtex4}
\usepackage{bbm}
\usepackage{graphicx}
\usepackage{amsmath}
\usepackage{amsfonts}
\begin{document}
\title{Large $N_c$ Expansion and the Parity Violating $\pi, N, \Delta$ Couplings}

\author{Shi-Lin Zhu}
\email{zhusl@phy.pku.edu.cn} \affiliation{Department of Physics
and State Key Laboratory of Nuclear Physics and Technology,\\
Peking University, Beijing 100871, China}

\begin{abstract}
In the limit of large $N_c$ we first consider the $N_c$ ordering
of the various parity violating $\pi, N, \Delta$ couplings. Then
we derive the relations among these couplings and consistency
relations from the stability of these couplings under the chiral
loop corrections with and without the mass splitting between $N$
and $\Delta$. Especially we find that $h_\Delta =-{3\over
\sqrt{5}}h_\pi$ in the large $N_c$ limit, which correctly
reproduces the relative sign and magnitude of the "DDH" values for
these PV couplings.
\end{abstract}


\maketitle

\section{Introduction}\label{sec}

The long range parity violating (PV) force between nucleons are
mediated by pions with one vertex being parity violating. These PV
$\pi, N, \Delta$ couplings play an important role in the various
hadronic PV experiments and polarized ${\vec e}p$ scattering
experiments \cite{ddh}. For example, these PV couplings may
contribute to the nucleon anapole moment \cite{R_A} and PV nuclear
forces \cite{z1,z2}. They may be classified into: the Yukawa
coupling $h_\pi, h_\Delta$, the vector couplings $h_V^{0,1,2}$ and
the axial vector couplings $h_A^{1,2}$ etc \cite{ks,R_A,h_pi}.

The Yukawa coupling $h_\pi$ was estimated using $SU(6)$ symmetry,
inputs from hyperon decay phenomenology and quark model
\cite{ddh,fcdh}. The one loop chiral corrections to $h_\pi$ up to
order $1/\Lambda_\chi^3$ were presented in \cite{h_pi} where the
corrections from $h_A^1, h_\Delta$ etc found to be significant.
Despite these efforts, both theoretical and experimental
information of these PV couplings is still rather scarce.

In this work we want to explore these PV coupling constants in the
framework of the large $N_c$ expansion. We first review the large
$N_c$ expansion formalism and collect some useful results from
literature in Section \ref{sec1}. We present the $N_c$ ordering of
$h_\pi, h_V^i, h_A^i$ in Section \ref{sec2}. Then we relate the
$\pi NN$ PV couplings to the $\pi \Delta\Delta$ and $\pi N\Delta$
couplings since the nucleon and delta resonance are in the same
band in the large $N_c$ limit. Several interesting relations are
found such as
\begin{equation}
h_\Delta =-{3\over \sqrt{5}}h_\pi\;.
\end{equation}
In Section \ref{sec4} we discuss the $N_c$ orderings of the chiral
loop corrections. The last section is a short summary.

\section{Large $N_c$ formalism}
\label{sec1}

Quantum Chromodynamics (QCD) is very complicated at the hadronic
scale. In order to explore the hadron structure, various
theoretical frameworks are proposed such as lattice QCD, the
chiral perturbation theory ($\chi$PT), QCD sum rule (QSR) etc. QSR
works at the typical hadronic scale. The expansion parameter of
$\chi$PT is $p/\Lambda_\chi$ where $p$ is the pion momentum. The
convergence of the chiral expansion series requires that the
typical momentum of the process be small. In contrast, the
expansion parameter of perturbative QCD is the strong coupling
constant $\alpha_s(Q^2)$ where $Q^2$ should be large.

t' Hooft first suggested an alternative expansion scheme in terms
of $N_c^{-1}$ with $N_c$ the color number \cite{tooft}. Such an
expansion is valid in the whole momentum range so long as $N_c$ is
large. Some salient features of $N_c=3$ theory (QCD) like
asymptotic freedom, confinement, chiral symmery and its
spontaneous breaking have to be kept if the large $N_c$ theory is
similar to QCD. The requirement that the large $N_c$ theory is a
nontrivial asymptotic one leads to the scaling behavior $g_s\sim
{\cal O}(N_c^{-{1\over 2}})$ from the gluon self energy diagram.

In the large $N_c$ limit the planar diagrams are dominant and the
internal quark loops are suppressed while the gluon loops are not.
Mesons are stable with its mass $\sim {\cal O}(1)$ and decay width
$\sim {\cal O}(N_c^{-1})$. The pion decay constant $F_\pi \sim
{\cal O}(\sqrt{N_c})$. A nice review of meson properties in the
large $N_c$ limit was presented in Ref. \cite{witten}. Witten
extended the $N_c$ counting rules to the baryons based on the mean
field theory picture \cite{witten}. We quote some results which we
need below. Baryon masses are $\sim {\cal O}(N_c)$ and the meson
baryon scattering amplitude is $\sim {\cal O}(1)$ at most.

\begin{figure*}
\begin{center}
\scalebox{0.8}{\rotatebox{-90}{\includegraphics{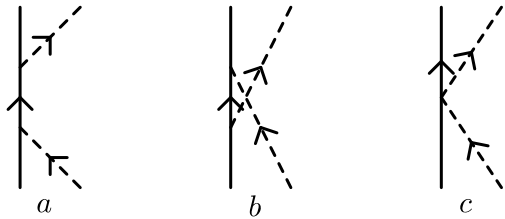}}}
\end{center}
\caption{Feyman diagrams for parity conserving pion baryons
scattering process. The solid and dashed line corresponds to the
nucleon and pion respectively. \label{fig1}}
\end{figure*}

\subsection{Operator formalism and large $N_c$ expansion}

Several groups considered the pion baryon scattering process in
Fig. \ref{fig1}(a)-(b) to derive the consistency conditions
\cite{gs,d1,dm}. [Fig. \ref{fig1}(c) is suppressed by $1/N_c$
compared to Fig. \ref{fig1}(a)-(b)]. The amplitude reads
\begin{eqnarray}\label{pp}
A\sim &{N_c^2 g^2_A\over F^2_\pi}
 {q_iq_j\over \omega} [X^{ia}, X^{jb}]
\end{eqnarray}
where $q_{i,j}$ is the initial and final pion three-momentum,
$g_A$ is the axial charge, $\omega$ is the pion energy, $X^{ia}$
is the operator for the pion baryon interaction vertex with
$I=J=1$, $i,a$ is the spin and isopsin indices respectively. A
factor $N_c$ has been extracted since pions can couple to $N_c$
quarks at each vertex. The coefficient in (\ref{pp}) is $\sim
{\cal O}(N_c)$. Unless the commutator vanishes the amplitude is
$\sim {\cal O}(N_c)$, which violates Witten's large $N_c$ counting
rules that the meson baryon scattering amplitude is $\sim {\cal
O}(1)$ at most. Naturally we have the consistency conditions:
\begin{eqnarray}\label{cd}
 [X^{ia}, X^{jb}]=0
\end{eqnarray}
in the leading order of $1/N_c$ expansion. However if the nucleon
is the only intermediate state, the above commutator definitely
does not hold. In other words, other states are needed to fulfill
Eq. (\ref{cd}). In fact the consistency condition requires the
presence of a band of states with $I=J=1/2, 3/2, \cdots$
\cite{d1,dashen}. From now on summation over all possible
intermediate states are assumed in the commutator.

For the phenomenological application it's convenient to use the
formalism of operator expansion and operator algebra generated by
$SU(4)$ spin flavor symmetry group \cite{georgi,d1}. In such an
approach the large $N_c$ counting of a specific process is made
explicit by relevant operators and their reduction for the low
lying baryons in the band. For the two flavor case, the
commutation relations for the generators of $SU(4)$ algebra are
\begin{eqnarray}\label{com}\nonumber
&[J^i, T^a]=0\; , \\ \nonumber &[J^i, J^j]=i\epsilon^{ijk} J^j \;
, \\ \nonumber &[T^a, T^b]=i\epsilon^{abc} T^c\; , \\ \nonumber
&[J^i, G^{ja}]=i\epsilon^{ijk} G^{ka} \; , \\ \nonumber
&[T^a, G^{ib}]=i\epsilon^{abc} G^{ic} \; , \\
&[G^{ia}, G^{jb}]={i\over 4}\delta^{ij}\epsilon^{abc} T^c+{i\over
4}\delta^{ab}\epsilon^{ijk} J^k
\end{eqnarray}
where $X^{ia}=G^{ia}/N_c$ \cite{d1,dashen}.

In the large $N_c$ limit the baryons form a band with $I=J=1/2,
3/2, \cdots$ \cite{d1,dashen}. For baryons lying in the upper part
of the band the matrix elements of the generators are naturally of
the order of $N_c$,
\begin{eqnarray}\label{poo}\nonumber
\langle N^\ast | T^a |N^\ast \rangle \sim {\cal O}(N_c)&\\
\nonumber
\langle N^\ast | J^i |N^\ast \rangle \sim {\cal O}(N_c)&\\
\langle N^\ast | G^{ia} |N^\ast \rangle \sim {\cal O}(N_c)
\end{eqnarray}
where we have denoted these highly lying states by $N^\ast$.
However these physically interesting states, which correspond to
the nucleon and delta baryon in the $N_c=3$ world, do lie at the
bottom of the band. In other words,
\begin{eqnarray}\label{po}\nonumber
\langle N, \Delta | T^a |N, \Delta \rangle \sim {\cal O}(1)&\\
\nonumber
\langle N, \Delta | J^i | N, \Delta\rangle \sim {\cal O}(1)&\\
\langle N, \Delta | G^{ia} | N, \Delta\rangle \sim {\cal O}(N_c)
\end{eqnarray}
For the two flavor case the spin and isospin of the nucleon and
delta baryon remains to be order of unity even in the large $N_c$
limit. This fact is very useful and important, which can be used
to expand operators in terms of $N_c$. For example the baryon mass
operator ${\hat H}$ must respect the rotation and isospin
symmetry. The lowest order few terms of ${\hat H}$ are
\cite{georgi,dashen,d1}
\begin{equation}\label{mass}
{\hat H}=N_c m_0 {\hat 1} +m_1 {J^2\over N_c}+ \cdots
\end{equation}
where we have used the fact that $I^2=J^2$ in the large $N_c$
limit for the two favor case. For those $N^*$ states every
operator in Eq. (\ref{mass}) contributes at the same $N_c$ order,
$\sim {\cal O} (N_c)$. If we focus on the nucleon and delta
resonances instead, we notice that their mass splitting is
\begin{equation}
{m_\Delta -m_N\over m_\Delta +m_N}\sim {1\over N_c^2}
\end{equation}
The above equation seems to hold even for the physical value
$N_c=3$.

Another example is the pion nucelon scattering amplitude in Fig.
\ref{fig1}(a)-(b).
\begin{eqnarray}\label{ga}\nonumber
A&\sim& {g^2_A\over F^2_\pi}
 {q_iq_j\over \omega} [G^{ia}, G^{jb}]\\ \nonumber
&=&{i\over 4}{g^2_A\over F^2_\pi} {q_iq_j\over \omega}
[\delta^{ij}\epsilon^{abc} T^c+\epsilon^{ijk} J^k] \\
&\sim & {\cal O}(N_c^{-1})
\end{eqnarray}
where we have used Eq. (\ref{com}) and (\ref{po}). However if we
take into account the mass splitting between the intermediate
state and initial and final states, we have \cite{j00}
\begin{eqnarray}\label{gb}\nonumber
A&\sim& {g^2_A\over F^2_\pi}
 {q_iq_j\over \omega^2} [G^{ia}, [{\hat M}, G^{jb}]]\\ \nonumber
&\sim& {g^2_A\over F^2_\pi}
 {q_iq_j\over \omega} [G^{ia}, [{J^2\over N_c}, G^{jb}]]\\ \nonumber
&\sim &{g^2_A\over F^2_\pi} {1\over N_c} \{G^{ia}, G^{jb}\} \\
&\sim & {\cal O}(1)
\end{eqnarray}
since $\{G^{ia}, G^{jb}\}\sim {\cal O}(N_c^2)$. A simpler way to
understand the above relation is to expand the baryon propagator
${i\over v\cdot k -\delta +i\epsilon}$ in Fig. \ref{fig1}(a) and
${i\over -v\cdot k -\delta +i\epsilon}$ in Fig. \ref{fig1}(b) to
the first order in $\delta$, the correction to the amplitude due
to the mass splitting reads,
\begin{eqnarray}\label{gc}\nonumber
A&\sim& {g^2_A\over F^2_\pi}
 {q_iq_j\over \omega^2} \delta \{G^{ia}, G^{jb}\}\\ \nonumber
&\sim &{1\over N^2_c} \{G^{ia}, G^{jb}\} \\
&\sim & {\cal O}(1)
\end{eqnarray}
where we have used $\delta \sim {\cal O}(N_c^{-1})$.

\subsection{Loop corrections in the large $N_c$ expansion}

Note the chiral loop corrections are not always suppressed in the
large $N_c$ counting. For example the dominant piece from the
chiral correction to the baryon mass is $\sim {\cal O}(N_c)$ and
does not break the degeneracy of the spectrum \cite{dm}. The
subleading symmetry breaking term from the loops is $\sim {\cal
O}(N_c^{-1})$. The correction from the mass splitting $\delta$
reads
\begin{equation}
A\sim {g_A^2\over F_\pi^2} \delta \{ G^{ia}, G^{ia}\} \sim {\cal
O}(1)
\end{equation}
Generally the correction from mass splitting is suppressed by
$1/N_c$ compared the leading term.

Consider a general operator $V^{\cdots}$ where $\cdots$ denotes
the isospin and spin indices. The one loop chiral correction
arises from the vertex and self energy diagrams \cite{j62}
\begin{equation}\label{vv}
\sim {N_c^2 g_A^2\over F_\pi^2} [X^{ia}, [X^{ia}, V^{\cdots}]]
\end{equation}
where the sum over the spin index arises after finishing the
momentum integral. The coefficient in Eq. (\ref{vv}) is $\sim
{\cal O}(N_c)$. The stability of the operator $V^{\cdots}$ under
the chiral loop corrections leads to the consistency condition
\begin{equation}\label{vv-1}
[X^{ia}, [X^{ia}, V^{\cdots}]]=0
\end{equation}
Eq. (\ref{vv-1}) is valid for any operator. If we include the mass
splitting, we have a correction term to (\ref{vv}) \cite{j62}
\begin{equation}\label{qw}
\sim {N_c^2 g_A^2\over F_\pi^2} \{ X^{ia}, [V^{\cdots}, [{\hat M},
X^{ia}]]\}
\end{equation}
From (\ref{qw}) we have the following consistency condition
\begin{equation}\label{qwr}
 \{ X^{ia}, [V^{\cdots}, [J^2, X^{ia}]]\}\sim {\cal O} (1)
\end{equation}

\section{Large $N_c$ counting for PV coupling constants}
\label{sec2}

\begin{figure*}
\begin{center}
\scalebox{0.8}{\rotatebox{-90}{\includegraphics{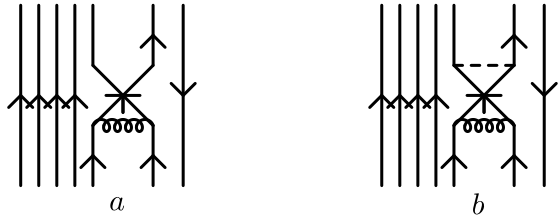}}}
\end{center}
\caption{The parity conserving (a) and violating (b) pion baryons
scattering process at the quark level. The dashed line in (b)
denotes the vector bosons $W^\pm, Z$. The curly line is the
gluon.\label{fig2} }
\end{figure*}

\begin{figure*}
\begin{center}
\scalebox{0.8}{\rotatebox{-90}{\includegraphics{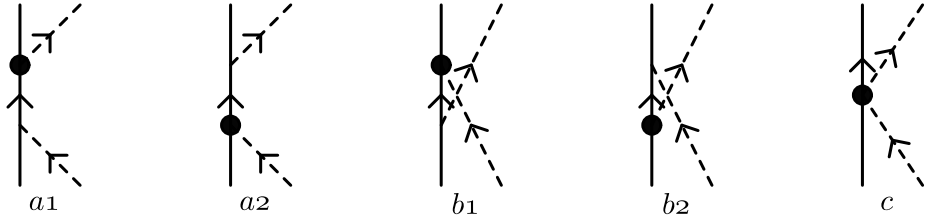}}}
\end{center}
\caption{Feynman diagrams for the parity violating pion baryons
scattering process. The filled circle denotes parity violating
vertex.  \label{fig3}}
\end{figure*}

Now we move on to the parity violating meson baryon scattering in
Fig. \ref{fig2}(b). Since the exchange of the W or Z bosons does
not change the color flow, the diagram Fig. \ref{fig2}(b) is of
the same order in the $N_c$ counting as Fig. \ref{fig2}(a) where
there is no vector boson exchange. From this observation we know
that the parity violating pion nucleon scattering process $\pi^a +
N \to \pi^b +N$ shown in Fig. \ref{fig3} is $\le {\cal O}(1)$ in
the $N_c$ counting and ${\cal O}(G_F)$.

The scattering amplitude of Fig. \ref{fig3}(c) is $\sim
{h_A^i\over F_\pi^2}\le {\cal O}(1)$. We have
\begin{equation}
h_A^i\le {\cal O} (N_c)\;.
\end{equation}
We will derive a more rigorous constraint below that
\begin{equation}
h_A^1\le  {\cal O} (1),
\end{equation}
\begin{equation}
h_A^2\le {\cal O} (N_c^{-1})
\end{equation}
from the consistency conditions for the chiral loop corrections in
Section \ref{sec4}.

Consider Fig. \ref{fig3} (a1)-(a2) and (b1)-(b2) with the PV
Yukawa insertions. The scattering amplitude at the leading order
of the $N_c$ counting is
\begin{eqnarray}\label{hpi}\nonumber
A&\sim &{h_\pi N_c g_A\over F_\pi}
\{ {q_i\over \omega} [X_0^{i-}, T^+]-{q'_i\over \omega} [X_0^{i+}, T^-]\}\\
&\sim &{h_\pi N_c g_A\over F_\pi} {q_i+q'_i\over \omega} X_0^{i0}
\end{eqnarray}
The commutator in the first line of (\ref{hpi}) sums all the
intermediate states. But the PV Yukawa interaction does not change
spin. Hence the intermediate state must have the same spin and
isospin as the initial or final state. So the commutator does not
vanish, which is strong contrast with the parity conserving pion
nucleon scattering where the summation is over the entire $I=J$
band. In the present case the summation is over a particular state
only due to the specific Lorentz structure of the PV Yukawa
vertex. Since $A\sim {\cal O}(1)$, we arrive at
\begin{equation}
h_\pi \sim {\cal O} ({1\over \sqrt{N_c}})\;.
\end{equation}

If we insert the PV vector coupling, only the time component is
relevant after the non-relativistic reduction. The discussion is
similar to (\ref{hpi}). We have
\begin{equation}
h_V^i \sim {\cal O}(1)\;.
\end{equation}

\section{The relations between PV coupling constants}
\label{sec3}

There exist several interesting relations between various PV
coupling constants in the large $N_c$ limit.

\medskip
\noindent (i) PV Yukawa coupling

In the $N_c\to \infty$ limit only baryons with $I=J=1/2, 3/2,
\cdots$ are relevant. The operator for the PV Yukawa pion baryon
interaction does not change the spin, i.e., $\Delta J=0$. The
initial and final baryons have the same spin and isopsin. That's
exactly what we have for the PV $\pi NN$ and $\pi \Delta\Delta$
Yukawa interaction. There does not exist the PV Yukawa $\pi
N\Delta$ coupling as required by the large $N_c$ argument. Let's
denote the PV Yukawa operator by $Y^{i=0,a}$ with the isospin
index $a=\pm$. The PV Yukawa couplings are determined by the
following matrix elements
\begin{eqnarray}
 \langle I_f,m_f;J_f,j_f | Y^{i=0,a} | I_i,m_i; J_i,j_i \rangle =
\sqrt{2I_i+1\over 2I_f +1} \left(
\begin{array}{lll}
I_i & 1& I_f\\
m_i& a &m_f
\end{array}
\right) \left(
\begin{array}{lll}
J_i & 0& J_f\\
j_i& 0 &j_f
\end{array}
\right) c_Y
\end{eqnarray}
where $I_{i,f}=J_{i,f}$ is the isospin (spin) of the initial and
final baryons, $m_i, j_i$ etc is the third component. $c_Y$ is a
constant which can not be determined with the $N_c$ argument only.

The tree level operator $Y^{i=0,a}$ in the PV Yukawa Lagrangians
is $\tau^\pm$. So for the PV $\pi NN$ Yukawa coupling we have
\begin{equation}\label{n-1}
h_\pi=c_Y \left(
\begin{array}{lll}
{1\over 2} & 1& {1\over 2}\\
-{1\over 2}& 1 & {1\over 2}
\end{array}
\right)=\sqrt{2\over 3}c_Y
\end{equation}
while for the PV $\pi \Delta\Delta$ Yukawa coupling we have
\begin{equation}\label{d-1}
{h_\Delta\over \sqrt{3}}=c_Y \left(
\begin{array}{lll}
{3\over 2} & 1& {3\over 2}\\
{1\over 2}& 1 & {3\over 2}
\end{array}
\right)=-\sqrt{2\over 5}c_Y
\end{equation}
where the factor ${1\over \sqrt{3}}$ arises from the DDH
convention for $h_\Delta$, which is a linear combination of
$h_{\pi\Delta}^1$ and $h_{\pi\Delta}^2$ in (\ref{ddd2}). Combining
(\ref{n-1}) and (\ref{d-1}) we have
\begin{equation}
h_\Delta =-{3\over \sqrt{5}} h_\pi
\end{equation}
It's very interesting to note $h_\Delta$ and $h_\pi$ has opposite
sign in the large $N_c$ limit, which is consistent with the DDH
phenomenological analysis based on $SU(6)$ symmetry and inputs
from hyperon decay data and quark model. The DDH ranges are $(0\to
17)g_\pi$ for $h_\pi$ and $(-51\to 0)g_\pi$ for $h_\Delta$ where
$g_\pi=3.8\times 10^{-8}$ \cite{ddh,fcdh}. Our large $N_c$
argument correctly reproduces the relative sign and magnitude.

\medskip
\noindent (ii) PV vector couplings

Simiar relations hold for the PV vector and axial vector pion
baryon couplings. For the vector PV coupling case, only the time
component of the vector operator $V^{i=0,a}$ remains after the
non-relativistic reduction, i.e., $\Delta J=0$. Therefore we do
not expect the vector-like $\pi N \Delta$ PV tree level
interaction from the large $N_c$ argument. Analysis in
\cite{R_A,h_pi} explicitly shows that such a PV coupling is
suppressed at the leading order of the heavy baryon expansion.

\medskip
\noindent (1) $\Delta I=0$ case:

Now let's move on to the PV coupling $h_V^0$ in (\ref{n1}) and
$j_0$ in (\ref{ddd1}) respectively. The relevant operator has spin
zero and isospin one.
\begin{equation}\label{n-2}
h_V^0=c_V^0 \left(
\begin{array}{lll}
{1\over 2} & 1& {1\over 2}\\
{1\over 2}& 0 & {1\over 2}
\end{array}
\right)=-\sqrt{1\over 3}c_V^0
\end{equation}
\begin{equation}\label{d-2}
j_0=c_V^0 \left(
\begin{array}{lll}
{3\over 2} & 1& {3\over 2}\\
{3\over 2}& 0 & {3\over 2}
\end{array}
\right)=\sqrt{3\over 5}c_V^0
\end{equation}
Hence $j_0=-{3\over \sqrt{5}}h_V^0$.

\medskip
\noindent (2) $\Delta I=1$ case:

For $h_V^1$ in (\ref{n2}) and $j_1$ in (\ref{ddd2}) the isospin
violation arises solely from the trace part $\mbox{Tr} (A_\mu
X_+^3)$. The relevant operator does not carry isospin or spin. So
we simply have $h_V^1 =j_1$. Note $j_{2,3,4}$ has no corresponding
terms in the PV $\pi NN$ Lagrangian.

\medskip
\noindent (3) $\Delta I=2$ case:

$h_V^2$ in (\ref{n3}) and $j_6$ in (\ref{ddd3}) involve $\Delta
I=2, \Delta I_z=0$. The relevant operator is ${\cal I}^{ab} A^a
\tau^b$ at the tree level. Hence, $j_6=-{3\over \sqrt{5}}h_V^2$.
In contrast, $j_5$ has not any similar term in the PV $\pi NN$
Lagrangian.

\medskip
\noindent (iii) PV axial vector couplings

Due to our vector spinor formalism for the $\Delta$ field the
operator for the PV axial vector $\pi N \Delta$ couplings in
(\ref{d2})-(\ref{d3}) have a rather different form as those in
(\ref{n2})-(\ref{n3}) and (\ref{ddd2})-(\ref{ddd3}). However after
the non-relativistic reduction, they have the same isospin and
spin structure since the nucleon and delta lie in the same
$I=J=1/2, 3/2, \cdots$ tower.

\medskip
\noindent (1) $\Delta I=1$ case:

For $h_A^1$ in (\ref{n2}) and $k_1$ in (\ref{ddd2}) the isospin
violation arises solely from $\mbox{Tr} (A_\mu X_-^3)$. The
relevant operator has spin one and isospin zero. Similarly we have
$k_1=-{3\over \sqrt{5}}h_A^1$. $k_{2,3,4}$ has no similar terms in
(\ref{n2}).

\medskip
\noindent (2) $\Delta I=2$ case:

$h_A^2$ in (\ref{n3}) and $k_5$ in (\ref{ddd3}) involves operator
with $I=J=1$ at the tree level. This operator is similar to
$X^{ia}$ associated with the strong pionic vertex. So we have,
\begin{equation}\label{n-3}
-{h_A^2\over 2}= \left(
\begin{array}{lll}
{1\over 2} & 1& {1\over 2}\\
{1\over 2}& 0 & {1\over 2}
\end{array}
\right)^2c_A^2={1\over 3}c_A^2
\end{equation}
\begin{equation}\label{d-3}
k_5= \left(
\begin{array}{lll}
{3\over 2} & 1& {3\over 2}\\
{3\over 2}& 0 & {3\over 2}
\end{array}
\right)^2c_A^2={3\over 5}c_A^2
\end{equation}
where the factor $-{1\over 2}$ is our convention in the definition
of $h_A^2$. Clearly,
\begin{equation}
{(\ref{d-3})\over (\ref{n-3})} ={g_{\pi^0
\Delta^{++}\Delta^{++}}\over g_{\pi^0 PP}}={g_1\over g_A}={9\over
5}
\end{equation}
in the large $N_c$ limit. Hence
\begin{equation}
h_A^2=-{10\over 9}k_5\;.
\end{equation}
The above analysis can be extended to the PV axial vector $\pi
N\Delta$ couplings.

\section{The chiral loops }
\label{sec4}

\begin{figure*}
\begin{center}
\scalebox{0.8}{\rotatebox{-90}{\includegraphics{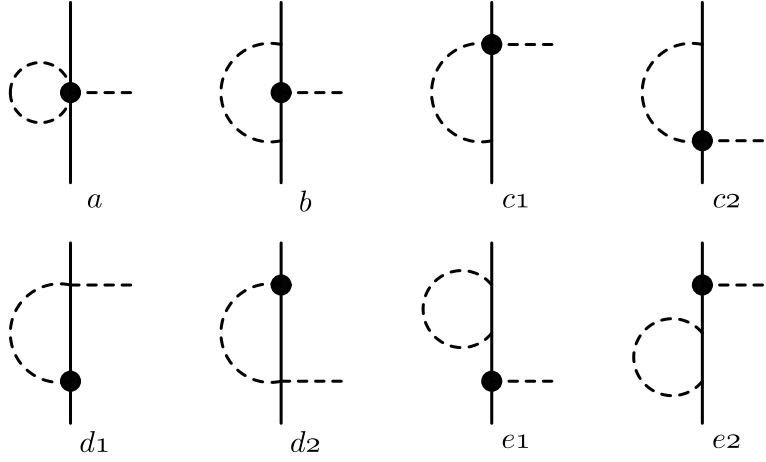}}}
\end{center}
\caption{The chiral loop corrections to $h_\pi$ from the nucleon
intermediate states. \label{fig4}}
\end{figure*}

\begin{figure*}
\begin{center}
\scalebox{0.8}{\rotatebox{-90}{\includegraphics{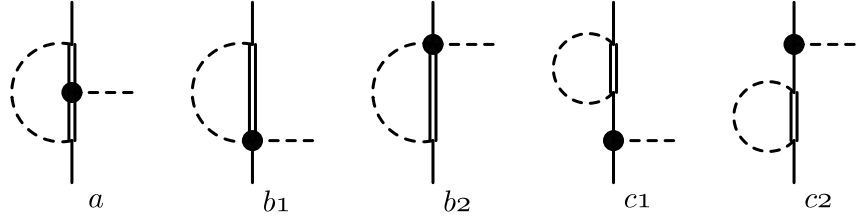}}}
\end{center}
\caption{Chiral loop corrections to $h_\pi$ from delta
intermediate states. The double line is the delta
resonance.\label{fig5} }
\end{figure*}

In this section we consider the large $N_c$ counting of the
various chiral loop corrections to the bare PV Yukawa coupling
$h_\pi$ in Fig. \ref{fig4}-\ref{fig5}, which was calculated in
\cite{h_pi}. The relevant operator for $h_\pi$ is $T^{\pm}$.

Fig. \ref{fig4}(a) arises from expanding the PV Yukawa $\pi NN$
vertex to the third order. Its contribution is
\begin{equation}
\sim {h_\pi\over F_\pi^2}\sim {\cal O}(N_c^{-1}) h_\pi\;
\end{equation}
which is suppressed by $1/N_c$ compared to the tree-level term
$h_\pi$.

The vertex correction and wave function renormalization
corrections in Fig. \ref{fig4}(b), \ref{fig4}(e1)-(e2) and Fig.
\ref{fig5}(a), \ref{fig5}(c1)-(c2) yield the sum
\begin{equation}\label{v}
\sim { g_A^2 h_\pi\over F_\pi^2} [G^{ia}, [G^{ia}, T^+]]\sim {
g_A^2 h_\pi\over F_\pi^2}T^+
\end{equation}
where we have used the commutators in (\ref{com}) twice. Clearly
its contribution is also $\sim {\cal O}(N_c^{-1}) h_\pi$. The
above relation holds in the limit of large $N_c$, i.e., when the
nucleon and delta states are degenerate in mass. If we explicitly
take into account the corrections of $\delta$ as done in
\cite{h_pi}, the loop correction reads
\begin{eqnarray}\label{v2}\nonumber
&\sim &{h_\pi g_A^2\over F_\pi^2}{1\over N_c} \{ G^{ia}, [T^+, [J^2, G^{ia}]]\}\\
&\sim &{\cal O}(1) h_\pi
\end{eqnarray}
which has the same $N_c$ order as the tree-level term! Fig.
\ref{fig4}(d1)-(d2) arises from the Yukawa vertex and expansion of
the chiral connection. Its contribution reads
\begin{equation}\label{w}
\sim { h_\pi\over F_\pi^2} [T^3, T^+]\sim {h_\pi\over F_\pi^2}T^+
\end{equation}
It is also $\sim {\cal O}(N_c^{-1}) h_\pi$. The correction from
mass splitting from these two diagrams is $\sim {\cal O}(N_c^{-2})
h_\pi$.

The contribution from the PV $\pi\pi NN$ and $\pi\pi N\Delta$
vertices are more subtle. This kind of contribution is very
similar to the chiral loop correction to the baryon mass. There
are two types of operators corresponding to the $\Delta I=1, 2$
two pions axial vertex respectively. In the leading order of the
large $N_c$ expansion, these operators can be read from Eq.
(\ref{n2})-(\ref{n3}). They are $J^i$ for the $h_A^1$ type and
$G^{ia}$ for the $h_A^2$ type. For the $h_A^1$ type, the sum of
Fig. \ref{fig4}(c1)-(c2) and Fig. \ref{fig5}(b1)-(b2) yields the
anti-commutator:
\begin{eqnarray}\label{a}\nonumber
&\sim& { g_A h_A^1 \over F_\pi^3} \{G^{ia}, J^i\} \\
&\sim & { g_A h_A^1 \over F_\pi^3}{N_c+2\over 2} T^a
\end{eqnarray}
where we have used the operator identity \cite{d1}
\begin{equation}
2\{G^{ia}, J^i\} =(N_c+2 )T^a
\end{equation}
From the stability of the Yukawa coupling $h_\pi$ under the chiral
correction, we conclude the loop correction is $\sim {\cal
O}(N_c^{-{1\over 2}})$ at most. Since $T^a \sim  {\cal O}(1)$,
$F_\pi \sim {\cal O}(\sqrt{N_c})$, we get
\begin{equation}
h_A^1 \le {\cal O}(1)\;.
\end{equation}
For the $h_A^2$ type correction, the sum of Fig.
\ref{fig4}(c1)-(c2) and Fig. \ref{fig5}(b1)-(b2) yields
\begin{eqnarray}\label{b}\nonumber
&\sim& { g_A h_A^2 \over F_\pi^3} G^{ia}G^{ib} \\ \nonumber
&\sim & { g_A h_A^2 \over 2F_\pi^3}\{[G^{ia}, G^{ib}]_-+ \{G^{ia}, G^{ib}\}_+\}\\
&\sim & N_c^{1\over 2} h_A^2
\end{eqnarray}
where we have used the identity \cite{d1}
\begin{equation}
\{G^{ia}, G^{ib}\} ={1\over 4}\{T^a, T^b\}+{\delta^{ab}\over
3}\{G^{ic}, G^{ic}\}_+ -{\delta^{ab}\over 4} T^2\; .
\end{equation}
From the same stability requirement we have
\begin{equation}
h_A^2 \le {\cal O}(N_c^{-1})\;.
\end{equation}

\section{Summary}
\label{sec5}

In short summary, we have investigated the $N_c$ ordering of the
parity violating $\pi, N, \Delta$ couplings in the limit of large
$N_c$. There exists several interesting relations between these
couplings. For the PV Yukawa type couplings $h_\Delta =-{3\over
\sqrt{5}}h_\pi$. Both the sign and magnitude are consistent with
the widely used "DDH" values for these PV coupling constants. The
stability of these couplings under the chiral loop corrections
leads to rather rigorous constraint on the $N_c$ ordering of the
PV axial vector type couplings. The loop correction of the PV
axial two pion vertex is of the same $N_c$ order as $h_\pi$, which
partly explains why $h_\pi$ receives a large radiative correction
from these couplings as noted in Ref. \cite{h_pi}.

\section*{Acknowledgments}

This project was supported by the National Natural Science
Foundation of China under Grants 10625521 and 10721063.

\newpage
\appendix
\section{PV $\pi NN$, $\pi \Delta\Delta$ and $\pi N\Delta$ Lagrangians}

We collect the PV pion nucleon delta Lagrangians below. Details
can be found in \cite{ks,R_A,h_pi}. The PV $\pi NN$ Lagrangian
reads
\begin{eqnarray}\label{n1}
{\cal L}^{\pi N}_{\Delta I=0} &=&h^0_V \bar N A_\mu \gamma^\mu N \\
\nonumber && \\
\label{n2} {\cal L}^{\pi N}_{\Delta I=1} &=&{h^1_V\over 2} \bar N
\gamma^\mu N  Tr (A_\mu X_+^3)
-{h^1_A\over 2} \bar N  \gamma^\mu \gamma_5N  Tr (A_\mu X_-^3)-{h_{\pi}\over 2\sqrt{2}}F_\pi \bar N X_-^3 N\\
\nonumber && \\
\label{n3} {\cal L}^{\pi N}_{\Delta I=2} &=&h^2_V {\cal I}^{ab}
\bar N [X_R^a A_\mu X_R^b +X_L^a A_\mu X_L^b]\gamma^\mu N
-{h^2_A\over 2} {\cal I}^{ab} \bar N [X_R^a A_\mu X_R^b -X_L^a
A_\mu X_L^b]\gamma^\mu\gamma_5 N \; ,
\end{eqnarray}

The analogues of Eqs. (\ref{n1}-\ref{n3}) for $\pi N\Delta$ are
\begin{eqnarray}\label{d1}
{\cal L}^{\pi\Delta N}_{\Delta I=0} &=&f_1 \epsilon^{abc} \bar N
i\gamma_5 [X_L^a A_\mu X_L^b +X_R^a A_\mu X_R^b] T_c^\mu  +g_1
\bar N [A_\mu, X_-^a]_+ T^\mu_a+g_2 \bar N [A_\mu, X_-^a]_-
T^\mu_a
+{\hbox{h.c.}}\\
\nonumber && \\
\label{d2} \nonumber {\cal L}^{\pi\Delta N}_{\Delta I=1} &=& f_2
\epsilon^{ab3} \bar N i\gamma_5 [A_\mu, X_+^a]_+ T^\mu_b +f_3
\epsilon^{ab3}\bar N i\gamma_5[A_\mu, X_+^a]_- T^\mu_b +{g_3\over
2}\bar N [(X_L^a A_\mu X_L^3-X_L^3 A_\mu X_L^a)\\
\nonumber && - (X_R^a A_\mu X_R^3-X_R^3 A_\mu X_R^a)]
T^\mu_a+{g_4\over 2} \{\bar N [3X_L^3 A^\mu (X_L^1 T^1_\mu +X_L^2
T^2_\mu ) + 3(X_L^1 A^\mu X_L^3 T^1_\mu \\
&& +X^2_L A^\mu X^3_L T^2_\mu) -2 (X_L^1 A^\mu X_L^1 +X_L^2 A^\mu
X_L^2-2X_L^3 A^\mu X_L^3)T^3_\mu] -(L\leftrightarrow R) \}
+{\hbox{h.c.}} \\
\nonumber && \\
\label{d3} \nonumber {\cal L}^{\pi\Delta N}_{\Delta I=2} &=& f_4
\epsilon^{abd} {\cal I}^{cd}\bar N i\gamma_5  [X_L^a A_\mu X_L^b
+X_R^a A_\mu X_R^b]T^\mu_c +f_5 \epsilon^{ab3} \bar N i\gamma_5
[X_L^a A_\mu X_L^3+X_L^3 A_\mu X_L^a \\
&& +(L\leftrightarrow R)]T^\mu_b +g_5 {\cal I}^{ab}\bar N [A_\mu,
X_-^a]_+ T^\mu_b +g_6 {\cal I}^{ab}\bar N [A_\mu, X_-^a]_- T^\mu_b
+{\hbox{h.c.}}\ \ \ ,
\end{eqnarray}

For the pv $\pi \Delta \Delta$ effective Lagrangians we have
\begin{equation}\label{ddd1}
{\cal L}^{\pi \Delta}_{\Delta I=0} =j_0 \bar T^i A_\mu \gamma^\mu
T_i \; ,
\end{equation}
\begin{eqnarray}\label{ddd2}\nonumber
{\cal L}^{\pi \Delta}_{\Delta I=1} ={j_1\over 2} \bar T^i
\gamma^\mu T_i Tr (A_\mu X_+^3) -{k_1\over 2} \bar T^i  \gamma^\mu
\gamma_5 T_i  Tr (A_\mu X_-^3) -{h^1_{\pi \Delta }\over
2\sqrt{2}}F_\pi \bar T^i X_-^3 T_i &\\ \nonumber -{h^2_{\pi \Delta
}\over 2\sqrt{2}}F_\pi \{ 3T^3 (X_-^1 T^1 +X_-^2 T^2) +3(\bar T^1
X_-^1 +\bar T^2 X_-^2 ) T^3  -2(\bar T^1 X_-^3 T^1 +\bar T^2 X_-^3
T^2 -2\bar T^3 X_-^3 T^3) \} &\\ \nonumber +j_2 \{ 3[(\bar T^3
\gamma^\mu T^1 +\bar T^1 \gamma^\mu T^3) Tr (A_\mu X_+^1) + (\bar
T^3 \gamma^\mu T^2 +\bar T^2 \gamma^\mu T^3) Tr (A_\mu X_+^2)] &\\
\nonumber -2(\bar T^1 \gamma^\mu T^1 +\bar T^2 \gamma^\mu T^2
-2\bar T^3 \gamma^\mu T^3 ) Tr (A_\mu X_+^3) \} &\\ \nonumber +k_2
\{ 3[(\bar T^3 \gamma^\mu\gamma_5 T^1 +\bar T^1 \gamma^\mu\gamma_5
T^3) Tr (A_\mu X_-^1) + (\bar T^3 \gamma^\mu\gamma_5 T^2 +\bar T^2
\gamma^\mu\gamma_5 T^3) Tr (A_\mu X_-^2)] &\\ \nonumber
 -2(\bar T^1 \gamma^\mu\gamma_5 T^1 +\bar T^2 \gamma^\mu\gamma_5 T^2 -2\bar T^3
\gamma^\mu\gamma_5 T^3 ) Tr (A_\mu X_-^3) \} &\\ \nonumber +j_3 \{
\bar T^a \gamma^\mu [A_\mu, X_+^a]_+ T^3 + \bar T^3 \gamma^\mu
[A_\mu, X_+^a]_+ T^a \}  +j_4 \{ \bar T^a \gamma^\mu [A_\mu,
X_+^a]_- T^3 - \bar T^3 \gamma^\mu [A_\mu, X_+^a]_- T^a \} &\\
 +k_3 \{ \bar T^a \gamma^\mu\gamma_5 [A_\mu, X_-^a]_+ T^3
+ \bar T^3 \gamma^\mu\gamma_5 [A_\mu, X_+^a]_+ T^a \} +k_4 \{ \bar
T^a \gamma^\mu\gamma_5 [A_\mu, X_-^a]_- T^3 - \bar T^3
\gamma^\mu\gamma_5 [A_\mu, X_+^a]_- T^a \}&\; ,
\end{eqnarray}
\begin{eqnarray}\label{ddd3}\nonumber
{\cal L}^{\pi \Delta}_{\Delta I=2} = j_5 {\cal I}^{ab} \bar T^a
\gamma^\mu A_\mu T^b +j_6 {\cal I}^{ab} \bar T^i [X_R^a A_\mu
X_R^b +X_L^a A_\mu X_L^b]\gamma^\mu T_i & \\  +k_5 {\cal I}^{ab}
\bar T^i [X_R^a A_\mu X_R^b -X_L^a A_\mu X_L^b]\gamma^\mu\gamma_5
T_i +k_6 \epsilon^{ab3} [\bar T^3 i\gamma_5 X_+^b T^a +\bar T^a
i\gamma_5 X_+^b T^3 ] &\; .
\end{eqnarray}

\end{document}